\documentclass{llncs}
\usepackage{array}
\usepackage{rotating}
\usepackage{multirow}
\usepackage{amsmath}
\usepackage{amssymb}
\usepackage{graphicx}
\usepackage{hyperref}
 \hypersetup{pdftitle={Distributed Verification of Rare Properties using Importance
Splitting Observers},pdfauthor={Sean Sedwards}}
\makeatletter

\providecommand{\tabularnewline}{\\}

 \usepackage[ruled,vlined]{algorithm2e}
 \usepackage{tikz}

\LinesNumbered
\let\oldnl\nl% Store \nl in \oldnl
\newcommand{\nonl}{\renewcommand{\nl}{\let\nl\oldnl}}% Remove line number for one line
\SetKwInput{KwIn}{input}
\SetKwInput{KwOut}{output}

\@ifundefined{showcaptionsetup}{}{%
 \PassOptionsToPackage{caption=false}{subfig}}
\usepackage{subfig}
\makeatother

\begin{document}

\title{Distributed Verification of Rare Properties using Importance Splitting
Observers}

\author{Cyrille Jegourel, Axel Legay, Sean Sedwards and Louis-Marie Traonouez}

\institute{INRIA Rennes -- Bretagne Atlantique}
\maketitle
\begin{abstract}
Rare properties remain a challenge for statistical model checking
(SMC) due to the quadratic scaling of variance with rarity. We address
this with a variance reduction framework based on lightweight importance
splitting observers. These expose the model-property automaton to
allow the construction of score functions for high performance algorithms.

The confidence intervals defined for importance splitting make it
appealing for SMC, but optimising its performance in the standard
way makes distribution inefficient. We show how it is possible to
achieve equivalently good results in less time by distributing simpler
algorithms. We first explore the challenges posed by importance splitting
and present an algorithm optimised for distribution. We then define
a specific bounded time logic that is compiled into memory-efficient
observers to monitor executions. Finally, we demonstrate our framework
on a number of challenging case studies.
\end{abstract}

\section{Introduction}

The `state explosion problem' \cite{ClarkeEmersonAllenSifakis2009}
associated with probabilistic model checking has been well addressed
by statistical model checking (SMC) \cite{YounesKwiatkowskaNormanParker2006}.
SMC includes a number of approximative techniques based on Monte Carlo
sampling \cite{MetropolisUlam1949}, which only generate states on
the fly during simulation. The performance of SMC is typically independent
of the size of the state space \cite{Niederreiter1992}, while simulation
cost may be divided linearly on parallel computation architectures.
Rare properties pose a problem because the standard and relative errors
scale quadratically with rarity \cite{HammersleyHandscomb1964,RubinoTuffin2009}.
For example, 4000 simulations would be sufficient to estimate a probability
of $0.1\pm10\%$ with $95\%$ confidence, whereas $4\times10^{13}$
simulations would be necessary to estimate a probability of $10^{-6}\pm10\%$
with the same confidence. Since quantifying rare properties is often
important to certify the reliability of complex critical systems,
we seek to enhance SMC with variance reduction techniques, such as
importance sampling and importance splitting \cite{KahnHarris1951,HammersleyHandscomb1964,RubinoTuffin2009},
while still taking advantage of the easy distribution that SMC typically
affords.

Importance sampling weights the executable model of a system so that
the rare property occurs more frequently in simulations. The proportion
of simulations that satisfy the property using the weighted model
overestimates the true probability, but the estimate may be exactly
compensated by the weights. It is generally not feasible to implement
a perfectly weighted executable model for importance sampling because
(\emph{i}) the perfect model may not actually exist as a re-parametrisation
of the original model and (\emph{ii}) a perfect re-parametrisation
typically requires an iteration over all the transitions, defeating
the benefits of sampling. Practical approaches tend to use a low dimensional
vector of parameters to weight the model \cite{SedwardsJegourelLegay2012,JegourelLegaySedwards2012}.
Given such a parametrisation, importance sampling can be implemented
with minimal memory and may be distributed efficiently on parallel
computational architectures. The principal limitation of importance
sampling is that without a guarantee that the simulation model is
perfect, it is difficult to formally bound the error of estimates.
In contrast, useful confidence intervals have been defined for importance
\emph{splitting} \cite{CerouGuyader2007,CerouDelMoralFuronGuyader2012}.

Importance splitting divides a rare property into a set of less rare
sub-properties that correspond to an increasing sequence of disjoint
levels: the initial state corresponds to the lowest level, while states
that satisfy the rare property corresponds to the final level. Importance
splitting algorithms use a series of easy simulation experiments to
estimate the conditional probabilities of going from one level to
the next. Since relatively few simulations fail to satisfy the sub-properties,
the overall simulation budget may be reduced. Each experiment comprises
simulations initialised with the terminal states of previous simulations
that reached the current level. The overall probability is the product
of the estimates, with the best performance (lowest variance) achieved
with many levels of equal conditional probability.

Importance splitting poses several challenges for optimisation and
distribution. In the context of SMC, importance splitting algorithms
repeatedly initialise simulations with states of the model-property
product automaton. For arbitrary properties this may have size proportional
to the length of a simulation trace. At the same time, increasing
the number of levels to maximise performance reduces the number of
simulation steps in each simulation experiment. The cost of sending
the model-property state across slow communication channels may be
significantly greater than the cost of short simulations. In addition,
to specify levels with equal conditional probabilities it is necessary
to define a `score function' that maps the states of the product
automaton to a value. This cannot easily be automated, so a syntactic
description of the property automaton must be accessible for the user
to construct a score function manually.

\smallskip{}

To address the above challenges we present an importance splitting
framework for SMC, specifically considering the problems of distribution.
We first discuss the problems of distributing importance splitting
algorithms and present a fixed level algorithm optimised for distribution.
We then define an expressive bounded time temporal logic and describe
the system of efficient lightweight observers that implement it. These
make the product automaton (\emph{i}) accessible to the user, (\emph{ii})
efficient to construct, (\emph{iii}) efficient to distribute and (\emph{iv})
efficient to execute. Finally, we demonstrate the performance and
flexibility of our framework on a number of case studies that are
intractable to numerical methods.

\subsection*{Related Work}

There have been many ad hoc implementations of importance splitting
based on the original ideas of \cite{KahnHarris1951,KahnMarshall1953}.
The algorithm of \cite{Villen-Altamirano1991} is a relatively recent
example that is often cited. The work of \cite{CerouGuyader2007,CerouDelMoralFuronGuyader2012}
is novel because the authors define efficient adaptive importance
splitting algorithms that also include confidence intervals. To our
knowledge, \cite{JegourelLegaySedwards2013} is the first work to
explicitly link importance splitting to arbitrary logical properties,
while the present work is the first to describe a practical importance
splitting framework for SMC. The present work is thus the first to
consider the problems of distributing importance splitting for SMC.

SMC tools construct an automaton (a monitor) to accept traces that
satisfy a temporal logic formula, typically based on a time bounded
variant of temporal logic. The proportion of independent simulations
of a stochastic model that satisfy the property is then used to estimate
the probability of the property or to test hypotheses about the probability.
There is considerable intersection between runtime verification (RV)
and SMC, with few concepts unique to either. In particular, there
have been many works that construct RV monitors from temporal logic
(e.g., \cite{Geilen2001,GiannakopoulouHavelund2001,HavelundRosu2002,FinkbeinerSipma2004,BauerLeuckerSchallhart2006}).
Such monitors typically comprise tableau-based automata \cite{GerthPeledVardiWolper1995}
whose states represent the combinations of subformulas of the overall
property. While some have considered timed properties (e.g., \cite{BauerLeuckerSchallhart2006}),
the focus is predominantly unbounded LTL properties interpreted on
finite paths \cite{Eisner-et-al2003}. In contrast, SMC typically
checks formulas with explicit time bounds (see, e.g., (\ref{eq:BLTL})),
which are inherently defined on finite traces. To avoid the combinatorial
explosion of subformulas caused by including time in this way, the
monitors used by \cite{SedwardsJegourelLegay2012,BoyerCorreLegaySedwards2013}
and other tools are compact ``programs'' that generate the states
of an automaton on the fly and do not store them. We adapt this ``lightweight''
approach to allow importance splitting for SMC to be efficiently distributed
on high performance parallel computational architectures.

\section{Technical Background}

Our SMC tools \cite{SedwardsJegourelLegay2012,BoyerCorreLegaySedwards2013}
implement a bounded linear temporal logic having the following typical
syntactic form:

\begin{equation}
\phi=\mathbf{X}^{k}\phi\mid\mathbf{F}^{k}\phi\mid\mathbf{G}^{k}\phi\mid\phi\mathbf{U}^{k}\phi\mid\neg\phi\mid\phi\vee\phi\mid\phi\wedge\phi\mid\phi\Rightarrow\phi\mid\alpha\label{eq:BLTL}
\end{equation}

This syntax allows arbitrary combinations and nesting of temporal
and atomic properties (i.e., those which may be evaluated in a single
state and denoted by $\alpha$). The time bound $k$ may denote discrete
steps or continuous time, but in this work we consider only discrete
time semantics.

Given a finite trace $\omega$, comprising sequence of states $\omega_{0}\omega_{1}\omega_{2}\cdots$,
$\omega^{(i)}$ denotes the suffix $\omega_{i}\omega_{i+1}\omega_{i+2}\cdots$.
The semantics of the satisfaction relation $\models$ is constructed
inductively as follows
\begin{equation}
\small\begin{split}\omega^{(i)}\models\, & true\\
\omega^{(i)}\models\, & \alpha\iff\alpha\textnormal{ is }\mathit{true}\textnormal{ in state }\omega_{i}\\
\omega^{(i)}\models\, & \neg\varphi\iff\omega^{(i)}\models\varphi\not\in\,\models\\
\omega^{(i)}\models\, & \varphi_{1}\vee\varphi_{2}\iff\omega^{(i)}\models\varphi_{1}\textnormal{ or }\omega^{(i)}\models\varphi_{2}\\
\omega^{(i)}\models\, & \mathbf{X}^{k}\varphi\iff\omega^{(k+i)}\models\varphi\\
\omega^{(i)}\models\, & \varphi_{1}\mathbf{U}^{k}\varphi_{2}\iff\exists j\in\{i,\dots,i+k\}:\omega^{(j)}\models\varphi_{2}\\
 & \wedge(j=i\vee\forall l\in\{i,\dots,j-1\}:\omega^{(l)}\models\varphi_{1})
\end{split}
\label{eq:semantics}
\end{equation}
Other elements of the relation are constructed using the equivalences
$\mathit{false}\equiv\neg\mathit{true}$, $\phi\wedge\phi\equiv\neg(\neg\phi\vee\neg\phi)$,
$\mathbf{F}^{k}\phi\equiv\mathit{true}\mathbf{U}^{k}\phi$, $\mathbf{G}^{k}\phi\equiv\neg(\mathit{true}\mathbf{U}^{k}\neg\phi)$.
Hence, given a property $\varphi$, with syntax according to (\ref{eq:BLTL}),
$\omega\models\varphi$ is evaluated by $\omega^{(0)}\models\varphi$.

\subsection*{Importance Splitting and Score Functions\label{sec:splitting}}

The neutron shield model of \cite{Kahn1950,KahnHarris1951} is illustrative
of how importance splitting works. The distance travelled by a neutron
in the shield defines a monotonic sequence of levels $0=s_{0}<s_{1}<s_{2}<\cdots<s_{m}=\mathit{shield\,thickness}$,
such that reaching a given level implies having reached all the lower
levels. While the overall probability $\gamma$ of passing through
the shield is small, the probability of passing from one level to
another can be made arbitrarily close to 1 by reducing the distance
between levels. Denoting the abstract level of a neutron as $s$,
the probability of a neutron reaching level $s_{i}$ can be expressed
as $\mathrm{P}(s\geq s_{i})=\mathrm{P}(s\geq s_{i}\mid s\geq s_{i-1})\mathrm{P}(s\geq s_{i-1})$.
Defining $\gamma=\mathrm{P}(s\geq s_{m})$ and $\mathrm{P}(s\geq s_{0})=1$,
\begin{equation}
\gamma=\prod_{i=1}^{m}\mathrm{P}(s\geq s_{i}\mid s\geq s_{i-1}).\label{eq:splitting}
\end{equation}
Each term of (\ref{eq:splitting}) is necessarily greater than or
equal to $\gamma$, making their estimation easier. By writing $\gamma_{i}=\mathrm{P}(s\geq s_{i}\mid s\geq s_{i-1})$
and denoting the estimates of $\gamma$ and $\gamma_{i}$ as respectively
$\hat{\gamma}$ and $\hat{\gamma}_{i}$, \cite{JegourelLegaySedwards2013}
defines the unbiased confidence interval 
\begin{equation}
\mathit{CI}=\left[\hat{\gamma}/\left(1+\frac{z_{\alpha}\sigma}{\sqrt{n}}\right),\hat{\gamma}/\left(1-\frac{z_{\alpha}\sigma}{\sqrt{n}}\right)\right]\qquad\textrm{with}\qquad\sigma^{2}\geq\sum_{i=1}^{m}\frac{1-\gamma_{i}}{\gamma_{i}}.\label{eq:confidence}
\end{equation}
Confidence is specified via $z_{\alpha}$, the $1-\alpha/2$ quantile
of the standard normal distribution, while $n$ is the per-level simulation
budget. We infer from (\ref{eq:confidence}) that for a given $\gamma$
the confidence is maximised by making both the number of levels $m$
and the simulation budget large, with all $\gamma_{i}$ equal.

The concept of levels can be generalised to arbitrary systems and
properties in the context of SMC, treating $s$ and $s_{i}$ in (\ref{eq:splitting})
as values of a score function over the model-property product automaton.
Intuitively, a score function discriminates good paths from bad, assigning
higher scores to paths that more nearly satisfy the overall property.
Since the choice of levels is crucial to the effectiveness of importance
splitting, various ways to construct score functions from a temporal
logic property are proposed in \cite{JegourelLegaySedwards2013}.
Formally, given a set of finite trace prefixes $\omega\in\Omega$,
an ideal score function $S:\Omega\rightarrow\mathbb{R}$ has the characteristics
$S(\omega)>S(\omega')\iff\mathrm{P}(\models\varphi\mid\omega)>\mathrm{P}(\models\varphi\mid\omega')$,
where $\mathrm{P}(\models\varphi\mid\omega)$ is the probability of
eventually satisfying $\varphi$ given prefix $\omega$. Intuitively,
$\omega$ has a higher score than $\omega'$ iff there is more chance
of satisfying $\varphi$ by continuing $\omega$ than by continuing
$\omega'$. The minimum requirement of a score function is $S(\omega)\geq s_{\varphi}\iff\omega\models\varphi$,
where $s_{\varphi}$ is an arbitrary value denoting that $\varphi$
is satisfied. Any trace that satisfies $\varphi$ must have a score
of at least $s_{\varphi}$ and any trace that does not satisfy $\varphi$
must have a score less than $s_{\varphi}$. In what follows we assume
that (\ref{eq:splitting}) refers to scores.

\section{Distributing Importance Splitting}

Simple Monte Carlo SMC may be efficiently distributed because once
initialised, simulations are executed independently and the result
is communicated at the end with just a single bit of information (i.e.,
whether the property was satisfied or not). By contrast, the simulations
of importance splitting are dependent because scores generated during
the course of each simulation must be processed centrally. The amount
of central processing can be minimised by reducing the number of levels,
but this generally reduces the overall performance.

Alternatively, entire instances of the importance splitting algorithm
may be distributed and their estimates averaged, with each instance
using a proportionally reduced simulation budget. We use this approach
to generate some of the results in Section \ref{sec:examples}, but
note that if the budget is reduced too far, the algorithm will fail
to pass from one level to the next and no valid estimate will be produced. 

Distribution of importance splitting is thus possible, but its efficiency
is dependent on the particular problem. In this work we therefore
provide the framework to explore different approaches. In Section
\ref{sec:adaptive} we first describe the concept of an adaptive importance
splitting algorithm and then explain why this otherwise optimised
technique is unsuitable for distribution. In Section \ref{sec:fixed}
we motivate the use of a fixed level algorithm for ``lightweight''
distribution and provide a suitable algorithm. The results we present
in Section \ref{sec:examples} demonstrate that this simpler approach
can be highly effective.

\subsection{The Adaptive Algorithm\label{sec:adaptive}}

The basic notion of importance splitting described in Section \ref{sec:splitting}
can be directly implemented in a so-called fixed level algorithm,
i.e., an algorithm in which the levels are pre-defined by the user.
With no a priori information, such levels will typically be chosen
to subdivide the maximum score equally. In general, however, this
will not equally divide the conditional probabilities of the levels,
as required by (\ref{eq:confidence}) to maximise performance. In
the worst case, one or more of the conditional probabilities will
be too low for the algorithm to pass between levels. Finding good
or even reasonable levels by trial and error may be computationally
expensive and has prompted the development of adaptive algorithms
that discover optimal levels on the fly \cite{CerouGuyader2007,JegourelLegaySedwards2013,JegourelLegaySedwards2014}.
Instead of pre-defining levels, the user specifies the proportion
of simulations to retain after each iteration. This proportion generally
defines all but the final conditional probability in (\ref{eq:splitting}).

The adaptive importance splitting algorithm first performs a number
of simulations until the overall property is decided, storing the
resulting traces of the model-property automaton. Each trace induces
a sequence of scores and a corresponding maximum score. The algorithm
finds a level that is less than or equal to the the maximum score
of the desired proportion of simulations to retain. The simulations
whose maximum score is below this current level are discarded. New
simulations to replace the discarded ones are initialised with states
corresponding to the current level, chosen at random from the retained
simulations. The new simulations are continued until the overall property
is decided and the procedure is repeated until a sufficient proportion
of simulations satisfy the overall property.

The principal advantage of the adaptive algorithm is that by simply
rejecting the minimum number of simulations at each level it is possible
to maximise confidence for a given score function. The principal disadvantage
is that it stores simulation traces, severely limiting the size of
model and simulation budget. The use of lightweight computational
threads is effectively prohibited. Moreover, minimising the number
of rejected simulations reduces the number of simulations performed
between levels, thus reducing the possibility to perform computations
in parallel. Minimising the rejected simulations also maximises the
number of levels, which in turn minimises the number of simulation
steps between each level. This further limits the feasibility of dividing
the algorithm, since sending a model-property state over a slow communication
channel may be orders of magnitude more costly than performing a short
simulation locally.

\subsection{A Fixed Level Algorithm for Distribution\label{sec:fixed}}

In contrast to the adaptive algorithm, the fixed level importance
splitting algorithm does not need to store traces, making it lightweight
and suitable for distribution. Scores are calculated on the fly and
only the states that achieve the desired level are retained for further
consideration. While the choice of levels remains a problem, an effective
strategy is to first use the adaptive algorithm with a relatively
high rejection rate to find good fixed levels. An estimate with high
confidence can then be generated efficiently by distributing the fixed
level algorithm.

Algorithm \ref{alg:fixed} is our fixed level importance splitting
algorithm optimised for distribution. We use the terms server and
client to refer to the root and leaf nodes of a network of computational
devices or to mean independent computational threads on the same machine.
In essence, the server manages the job and the clients perform the
simulations. The server initially sends compact representations of
the model and property to each client. Thereafter, only the state
of the product automaton is communicated. In general, each client
returns terminal states of simulations that reached the current level
and the server distributes these as initial states for the next round
of simulations. Algorithm \ref{alg:fixed} optimises this. The server
requests and distributes only the number of states necessary to restart
the simulations that failed to reach the current level, while maintaining
the randomness of the selection. Despite this optimisation, however,
the performance of this and other importance splitting algorithms
will be confounded by the combination of large state size and properties
having short time bounds. Under such circumstances it may be preferable
to distribute entire instances of the algorithm, as described above. 

The memory requirements of Algorithm \ref{alg:fixed} are minimal.
Each client need only store the state of $n$ simulations. As such,
it is conceivable to distribute simulations on lightweight computational
threads, such as those provided by GPGPU (general purpose computing
on graphics processing units).

\begin{algorithm}
\KwIn{$s_{1}<s_{2}<\cdots<s_{m}$ is a sequence of scores, with $s_{m}=s_{\varphi}$
the score necessary to satisfy property $\varphi$}

$\hat{\gamma}\leftarrow1$ is the initial estimate of $\gamma=\mathrm{P}(\omega\models\varphi)$

server sends compact description of model and observer to $k$ clients

each client initialises $n$ simulations

\For{$s\leftarrow s_{1},\dots,s_{m}$}{

each client continues its $n$ simulations from their current state\\\nonl{}simulations
halt as soon as their scores reach $s$

$\forall$ clients, client $i$ sends server the number of traces
$n_{i}$ that reached $s$

server calculates $\hat{\gamma}\leftarrow\hat{\gamma}n'/kn$, where
$n'=\sum n_{i}$

\For{$j\leftarrow1,\dots,kn-n'$}{

server chooses client $i$ at random, with probability $n_{i}/n'$

client $i$ sends server a state chosen uniformly at random from those
that reached $s$

server sends state to client corresponding to failed simulation $j$,
as initial state of new simulation to replace simulation $j$

}}\KwOut{$\hat{\gamma}$}

\protect\caption{Distributed Fixed Level Importance Splitting\label{alg:fixed}}
\end{algorithm}

\section{Linear Temporal Logic for Importance Splitting\label{sec:ISPLTL}}

High performance SMC tools, such as \cite{SedwardsJegourelLegay2012,BoyerCorreLegaySedwards2013},
avoid the complexity of standard model checking by compiling the property
to a program of size proportional to the formula and memory proportional
to the maximum sum of nested time bounds. This program implicitly
encodes the model checking automaton, but is exponentially smaller.
For example, the property $\mathbf{X}^{k}\phi$ can be implemented
as a loop that generates $k$ simulation steps before returning the
truth of $\phi$ in the last state; the property $\psi\mathbf{U}^{k}\phi$
can be implemented as a loop that generates up to $k$ simulation
steps while $\psi$ is true and $\phi$ is not true, returning the
value of $\phi$ in the last state otherwise. If $\psi$ and $\phi$
are atomic, the programs require just $\mathcal{O}(\log k)$ bits
of memory to hold a loop counter.

In contrast, the nested property $\mathbf{F}^{k1}(\psi\vee\mathbf{G}^{k2}\phi)$
has an $\mathcal{O}(k2)$ memory requirement. If $\psi$ is not true
on step $i<k1$ it may be necessary to simulate up to step $i+k2$
to decide subformula $\mathbf{G}^{k2}\phi$. If $\psi\vee\mathbf{G}^{k2}\phi$
turns out to be false on step $i$, it will then be necessary to consider
the truth of $\psi$ on step $i+1$, noting that the last simulated
step could be $i+k2$. To evaluate this formula it is effectively
necessary to remember the truth of $\psi$ on $\mathcal{O}(k2)$ simulation
steps. Similar requirements can arise when the until operator ($\mathbf{U}$)
is a subformula of a temporal operator. In all such cases, the sequence
of stored truth values become part of the state of the property automaton.

SMC using importance splitting requires that simulations are repeatedly
and frequently initialised with the state of the model-property product
automaton. If the size of this state is proportional to the time bounds
of temporal operators, initialisation may have comparable complexity
to simulation. This becomes especially problematic if the state is
to be transmitted across relatively slow communication channels for
the purposes of distribution. We therefore define a subset of (\ref{eq:BLTL}),
the size of whose automata is not dependent on the bounds of temporal
operators:
\begin{equation}
\begin{aligned}\begin{split}\phi= & \mathbf{X}^{k}\phi\mid\psi\mathbf{U}^{k}\psi\mid\neg\phi\mid\phi\vee\phi\mid\phi\wedge\phi\mid\phi\Rightarrow\phi\mid\psi\\
\psi= & \mathbf{X}^{k}\psi\mid\mathbf{F}^{k}\psi\mid\mathbf{G}^{k}\psi\mid\alpha
\end{split}
\end{aligned}
\label{eq:ISPLTL}
\end{equation}

The semantics of (\ref{eq:ISPLTL}) is the same as (\ref{eq:BLTL}),
but (\ref{eq:ISPLTL}) restricts how temporal operators may be combined.
In particular, $\mathbf{U}$ may not be the subformula of a temporal
operator other than $\mathbf{X}$ and temporal operators that are
subformulas of other temporal operators may not be combined with Boolean
connectives. Temporal operators containing other temporal operators
as subformulas may, however, be combined. This logic expresses many
useful properties, including nested bounded temporal properties that
are not implemented in \textsc{Prism}.

\section{Lightweight Observers for Importance Splitting\label{sec:observers}}

To facilitate the construction of score functions we implement the
logic given by (\ref{eq:ISPLTL}) as a set of nested observers. Each
observer corresponds to either a temporal operator, a Boolean operator
acting on temporal operators, or as a predicate describing an atomic
property. In our implementation observers are written in a syntax
based on the commonly used reactive modules language \cite{AlurHenzinger1999},
using the notion of `guarded commands' \cite{Dijkstra1975} with
sequential semantics.

An observer comprises a set of guarded commands, any number of which
may be enabled and executed on a given simulation step. Updates are
performed in syntactic order after all guards have been evaluated,
hence the update of one command does not affect the guards of commands
in the same observer. In general, the output of one observer is the
input to another and observers are therefore executed in reverse order
of their nesting.

Observers evaluate states as they are generated by the simulation.
Since it may not be possible to decide a property before seeing a
certain number of states, observers implement a three valued logic.
In Figs. \ref{fig:connectives}, \ref{fig:inner} and \ref{fig:Uouter}
we use the symbols \textsf{?}, $\top$ and $\bot$ to denote the three
values \emph{undecided}, \emph{true} and \emph{false}, respectively.
The state of an observer changes only when at least on of its inputs
is decided. An observer may reach a deadlock state (no commands enabled)
once its output is decided and cannot be changed by further input.
A simulation terminates when the output of the root observer is decided,
i.e., the property is decided. Simulations may also be paused by the
importance splitting algorithm if the score reaches a desired level.

Observers implementing the same temporal operator behave differently
according to their level of nesting within a formula. We therefore
distinguish \emph{outer} and \emph{inner} temporal observers. The
temporal operators closest to the root of any branch of the syntax
tree induced by a formula are implemented by outer observers. Their
output proceeds from \emph{undecided} to either \emph{true} or \emph{false}
and then does not change. Inner observers encode temporal operators
that are the subformulas of other temporal operators. Their output
proceeds from \emph{undecided} to a possibly alternating sequence
of \emph{true}, \emph{false} and \emph{undecided} values because their
enclosing operator(s) cause them to evaluate a moving widow of states.
The inner and outer variants of $\mathbf{X}$, $\mathbf{F}$ and $\mathbf{G}$
are closely related---outer observers are essentially simplified inner
observers. When $\mathbf{U}$ is a subformula of $\mathbf{X}$, however,
the $\mathbf{X}$ is implemented as a delay within the $\mathbf{U}$
observer.

In what follows we describe the important aspects of the various observers
that implement (\ref{eq:ISPLTL}). The accompanying figures include
a diagrammatic representation of how the observers work and a set
of commands written in the form $\mathit{predicate}:\mathit{update}$.
Each observer has Boolean output variables $o$ and $d$ to indicate
respectively the result and whether the property has been decided
(observers for atomic formulas omit $d$). Observers for temporal
operators take discrete time bound $k$ as a parameter and use a counter
variable $w$ ($\mathbf{U}$ uses counter variables $w'$ and $w''$).
Inner temporal operators make use of an additional counter, $t$ ($\mathbf{U}$
uses $t'$ and $t''$). The inputs of observers are Boolean variables
$o'$ and $o''$, with corresponding decidedness $d'$ and $d''$.

\subsubsection*{Connective Observers}

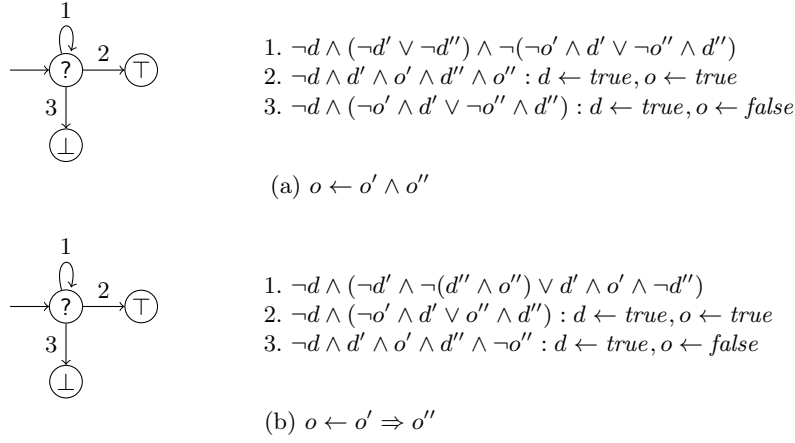
\begin{figure}
\subfloat[$o\leftarrow o'\wedge o''$\label{fig:outerconj}]{%
\begin{minipage}[c]{0.4\columnwidth}%
\centering
\begin{tikzpicture}
\tikzstyle{every node}=[draw,circle,inner sep=1pt]
\draw (-0.8,0)node[draw=none](I){}(0,0)node(0){\textsf{\,?\,}}(1,0)node(1){$\top$}(0,-1)node(2){$\bot$};
\tikzstyle{every node}=[draw=none]
\draw
(0)edge[loop above]node[above]{1}(0);
\draw[->]
(I)edge(0)
(0)edgenode[above]{2}(1)
(0)edgenode[left]{3}(2);
\end{tikzpicture}%
\end{minipage}%
\begin{minipage}[c]{0.58\columnwidth}%
1. $\neg d\wedge(\neg d'\vee\neg d'')\wedge\neg(\neg o'\wedge d'\vee\neg o''\wedge d'')$

2. $\neg d\wedge d'\wedge o'\wedge d''\wedge o'':d\leftarrow\mathit{true},o\leftarrow\mathit{true}$

3. $\neg d\wedge(\neg o'\wedge d'\vee\neg o''\wedge d''):d\leftarrow\mathit{true},o\leftarrow\mathit{false}$%
\end{minipage}

}

\subfloat[$o\leftarrow o'\Rightarrow o''$\label{fig:outerimpl}]{%
\begin{minipage}[c]{0.4\columnwidth}%
\centering
\begin{tikzpicture}
\tikzstyle{every node}=[draw,circle,inner sep=1pt]
\draw (-0.8,0)node[draw=none](I){}(0,0)node(0){\textsf{\,?\,}}(1,0)node(1){$\top$}(0,-1)node(2){$\bot$};
\tikzstyle{every node}=[draw=none]
\draw
(0)edge[loop above]node[above]{1}(0);
\draw[->]
(I)edge(0)
(0)edgenode[above]{2}(1)
(0)edgenode[left]{3}(2);
\end{tikzpicture}%
\end{minipage}%
\begin{minipage}[c]{0.58\columnwidth}%
1. $\neg d\wedge(\neg d'\wedge\neg(d''\wedge o'')\vee d'\wedge o'\wedge\neg d'')$

2. $\neg d\wedge(\neg o'\wedge d'\vee o''\wedge d''):d\leftarrow\mathit{true},o\leftarrow\mathit{true}$

3. $\neg d\wedge d'\wedge o'\wedge d''\wedge\neg o'':d\leftarrow\mathit{true},o\leftarrow\mathit{false}$%
\end{minipage}

}

\protect\caption{Connective observers. Initially $d=\mathit{false}$.\label{fig:connectives}}
\end{figure}

These observers implement Boolean connectives at syntactic level $\phi$
in (\ref{eq:ISPLTL}) and take advantage of the equivalences $\mathit{false}\wedge\textsf{?}=\mathit{false}$,
$\mathit{true}\vee\textsf{?}=\mathit{true}$, $\mathit{false}\Rightarrow\textsf{?}=\mathit{false}$
and $\textsf{?}\Rightarrow\mathit{true}=\mathit{true}$, for any truth
value of $\textsf{?}$. Figure \ref{fig:outerconj} describes the
observer for conjunction and Fig. \ref{fig:outerimpl} describes the
observer for implication. The observer for disjunction may be derived
from that of conjunction by negating all instances of $o'$ and $o''$,
and by exchanging $o\leftarrow\mathit{true}$ and $o\leftarrow\mathit{false}$.
Negation is implemented by inverting the truth assignment of the observer
to which it applies, i.e., by exchanging $o\leftarrow\mathit{true}$
and $o\leftarrow\mathit{false}$. The connectives may be combined
with themselves and outer temporal operators. Boolean connectives
that apply only to atomic properties (i.e., syntactic level $\alpha$)
are implemented directly in formulas within observers for atomic properties.

\subsubsection*{Inner Temporal Observers}

These observers act on a moving window of states created by an enclosing
temporal operator. The output may pass from one decided value to the
other and also become undecided.

\begin{figure}

\subfloat[$o\leftarrow\mathbf{X}^{k}o'$\label{fig:Xinner}]{%
\begin{minipage}[c]{0.4\columnwidth}%
\centering
\begin{tikzpicture}
\tikzstyle{every node}=[draw,circle,inner sep=1pt]
\draw (-0.8,0)node[draw=none](I){}(0,0)node(0){\textsf{\,?\,}}(1.4,0)node(1){$\top$}(0,-1.4)node(2){$\bot$};
\tikzstyle{every node}=[draw=none]
\draw
(0)edge[loop above]node[above]{1}(0)
(1)edge[loop right]node[right]{2}(1)
(2)edge[loop left]node[left]{2}(2);
\draw[->]
(I)edge(0)
(0)edge[bend left=15]node[above]{2}(1)
(1)edge[bend left=15]node[below]{3}(0)
(0)edge[bend right=15]node[left]{2}(2)
(2)edge[bend right=15]node[right]{3}(0)
(1)edge[bend left=20]node[left]{2\,~}(2)
(2)edge[bend right=40]node[right]{\,2}(1);
\end{tikzpicture}%
\end{minipage}%
\begin{minipage}[c]{0.58\columnwidth}%
1. $\neg d\wedge d'\wedge w<k:w\leftarrow w+1$

2. $d'\wedge w=k:d\leftarrow\mathit{true},o\leftarrow o'$

3. $d\wedge\neg d':d\leftarrow\mathit{false}$%
\end{minipage}

}

\subfloat[$o\leftarrow\mathbf{F}^{k}o'$\label{fig:Finner}]{%
\begin{minipage}[c]{0.4\columnwidth}%
\centering
\begin{tikzpicture}
\tikzstyle{every node}=[draw,circle,inner sep=1pt]
\draw (-0.8,0)node[draw=none](I){}(0,0)node(0){\textsf{\,?\,}}(1.4,0)node(1){$\top$}(0,-1.4)node(2){$\bot$};
\tikzstyle{every node}=[draw=none]
\draw
(0)edge[loop above]node[above]{1}(0)
(1)edge[loop right]node[right]{2,5}(1)
(2)edge[loop left]node[left]{3}(2);
\draw[->]
(I)edge(0)
(0)edge[bend left=15]node[above]{2}(1)
(1)edge[bend left=15]node[below]{4,6}(0)
(0)edge[bend right=15]node[left]{3}(2)
(2)edge[bend right=15]node[right]{6}(0)
(1)edge[bend left=20]node[left]{3\,}(2)
(2)edge[bend right=40]node[right]{\,2}(1);
\end{tikzpicture}%
\end{minipage}%
\begin{minipage}[c]{0.58\columnwidth}%
1. $\neg d\wedge d'\wedge\neg o'\wedge w<k:w\leftarrow w+1$

2. $d'\wedge o':o\leftarrow\mathit{true},d\leftarrow\mathit{true},t\leftarrow w$

3. $d'\wedge\neg o'\wedge t=0\wedge w=k:d\leftarrow\mathit{true},o\leftarrow\mathit{false}$

4. $d\wedge d'\wedge\neg o'\wedge t=0\wedge w<k:d\leftarrow\mathit{false}$

5. $d\wedge d'\wedge\neg o'\wedge t>0:t\leftarrow t-1$

6. $d\wedge\neg d':d\leftarrow\mathit{false}$%
\end{minipage}

}

\protect\caption{Observers for inner temporal operators. Initially $w=t=0,d=\mathit{false}$.\label{fig:inner}}
\end{figure}
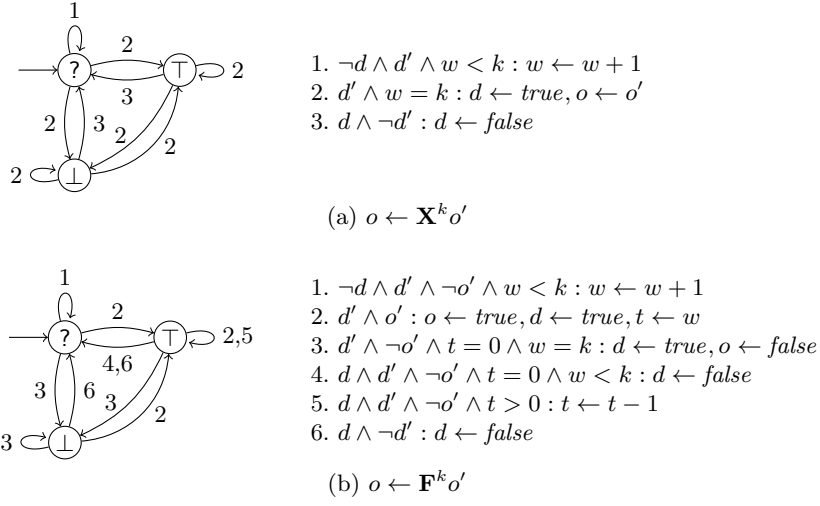

Figure \ref{fig:Xinner} describes the observer for $\mathbf{X}^{k}$.
Command 1 counts decided input states until bound $k$ is reached.
Thereafter command 2 sets the output decided and equal to the value
of the input.

Figure \ref{fig:Finner} describes the observer for $\mathbf{F}^{k}$.
While decided inputs are not \emph{true}, command 1 increments $w$
from 0 to $k$. If at any time the input is \emph{true}, command 2
sets the output to \emph{true} and the ``true-counter'' $t$ is
set to $w$. Command 5 decrements $t$ on subsequent false inputs.
The output remains true while $t>0$. If $w$ reaches $k$ while $t=0$,
command 3 sets the output to \emph{false}.

The observer for $\mathbf{G}^{k}$ may be derived from that of $\mathbf{F}^{k}$
by negating all instances of $o'$ and $\neg o'$, and by exchanging
$o\leftarrow\mathit{true}$ and $o\leftarrow\mathit{false}$.

\subsubsection*{Outer Temporal observers}

The outer observers for $\mathbf{X}^{k}$ and $\mathbf{F}^{k}$ are
not illustrated but may be derived from their respective inner observers
given in Fig. \ref{fig:inner}. For $\mathbf{X}^{k}$, command 3 is
removed and the guard of command 2 is strengthened with $\neg d$.
For $\mathbf{F}^{k}$, commands 4, 5 and 6, together with all references
to counter $t$, are removed, while the guards of commands 2 and 3
are strengthened by $\neg d$. The outer observer for $\mathbf{G}^{k}$
can be derived from that of $\mathbf{F}^{k}$ in the same way as described
for inner temporal observers.

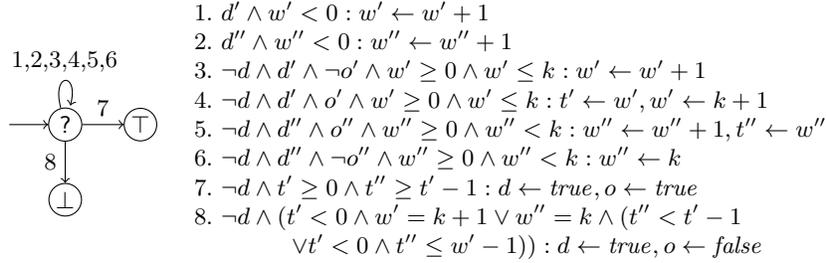
\begin{figure}
\begin{minipage}[c]{0.25\columnwidth}%
\centering
\begin{tikzpicture}
\tikzstyle{every node}=[draw,circle,inner sep=1pt]
\draw (-0.8,0)node[draw=none](I){}(0,0)node(0){\textsf{\,?\,}}(1,0)node(1){$\top$}(0,-1)node(2){$\bot$};
\tikzstyle{every node}=[draw=none]
\draw
(0)edge[loop above]node[above]{1,2,3,4,5,6}(0);
\draw[->]
(I)edge(0)
(0)edgenode[above]{7}(1)
(0)edgenode[left]{8}(2);
\end{tikzpicture}%
\end{minipage}%
\begin{minipage}[c]{0.74\columnwidth}%
1. $d'\wedge w'<0:w'\leftarrow w'+1$

2. $d''\wedge w''<0:w''\leftarrow w''+1$

3. $\neg d\wedge d'\wedge\neg o'\wedge w'\geq0\wedge w'\leq k:w'\leftarrow w'+1$

4. $\neg d\wedge d'\wedge o'\wedge w'\geq0\wedge w'\leq k:t'\leftarrow w',w'\leftarrow k+1$

5. $\neg d\wedge d''\wedge o''\wedge w''\geq0\wedge w''<k:w''\leftarrow w''+1,t''\leftarrow w''$

6. $\neg d\wedge d''\wedge\neg o''\wedge w''\geq0\wedge w''<k:w''\leftarrow k$

7. $\neg d\wedge t'\geq0\wedge t''\geq t'-1:d\leftarrow\mathit{true},o\leftarrow\mathit{true}$

8. $\neg d\wedge(t'<0\wedge w'=k+1\vee w''=k\wedge(t''<t'-1$

\qquad{}\qquad{}$\vee t'<0\wedge t''\leq w'-1)):d\leftarrow\mathit{true},o\leftarrow\mathit{false}$%
\end{minipage}

\protect\caption{Observer for $o\leftarrow\mathbf{X}^{k_{\mathbf{X}}}(o''\mathbf{U}^{k}o')$.
Initially $t''=0,t'=-1,d=d_{\mathbf{X}}=\mathit{false}$ and $w'=w''=-k_{\mathbf{X}}$
(see text).\label{fig:Uouter}}
\end{figure}

Figure \ref{fig:Uouter} describes the observer for properties of
the form $\mathbf{X}^{k_{\mathbf{X}}}(\psi\mathbf{U}^{k}\phi)$ and
$\psi\mathbf{U}^{k}\phi$. Since $\psi$ and $\phi$ may be temporal
formulas that are satisfied on different simulation steps in arbitrary
order, the observer employs variables $w'$ and $w''$ to respectively
count the sequences of $\neg\phi$ and $\psi$ (commands 3 and 5).
Variable $t'$ then records the position of the first $\phi$ (command
4), while $t''$ records the position of the last $\psi$ (command
5). Using $t'$ and $t''$, commands 7 and 8 are able to determine
if the property is satisfied or falsified, respectively.

Properties of the form $\mathbf{X}^{k_{\mathbf{X}}}(\psi\mathbf{U}^{k}\phi)$
are implemented by simply initialising variables $w'$ and $w''$
to $-k_{\mathbf{X}}$, forcing the observer to ignore the first $k_{\mathbf{X}}$
decided values of $\psi$ and $\phi$. If the property is not of this
form, $w'$ and $w''$ are initialised to 0 and the automaton may
be simplified by removing commands 1 and 2 and all instances of expressions
$w'\geq0$ and $w''\geq0$.

\vspace{-1em}

\section{Case Studies\label{sec:examples}}

We have implemented our importance splitting framework in \textsc{Plasma}
\cite{BoyerCorreLegaySedwards2013} and demonstrate its use on three
case studies whose state space is intractable to numerical model checking.
The following results do not seek to promote a particular methodology
(adaptive or fixed level algorithm, distributed or single machine),
but serve to illustrate the flexibility of our platform. The software,
models and observers can be downloaded from our website\textsc{}\footnote{projects.inria.fr/plasma-lab/importance-splitting\label{fn:webpage}}.
The leader election and dining philosophers models are also illustrated
on the \textsc{Prism} case studies website\footnote{www.prismmodelchecker.org/casestudies\label{fn:prism}}.

\begin{figure}
\begin{minipage}[t]{0.48\columnwidth}%
\includegraphics[width=0.9\textwidth]{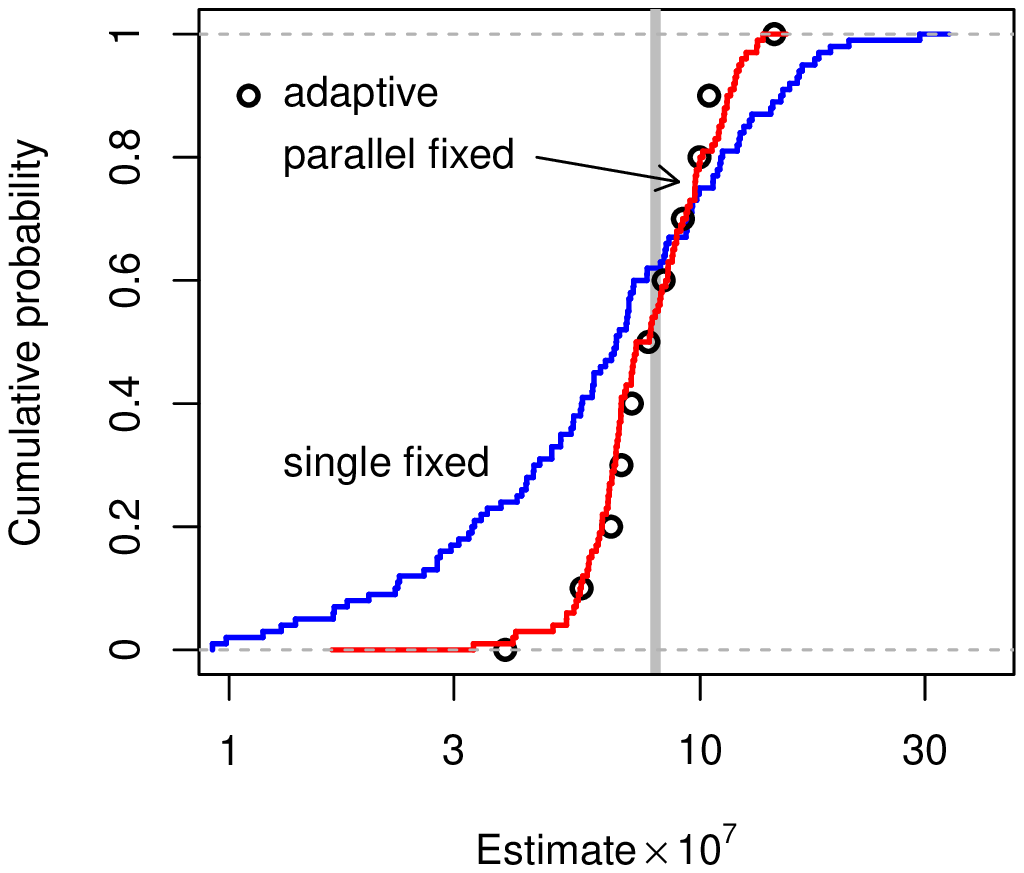}

\protect\caption{Dependent counters.\label{fig:counters}}
\end{minipage}\quad{}%
\begin{minipage}[t]{0.5\columnwidth}%
\textsf{\scriptsize{}}%
\begin{tabular}[b]{r|rccc|}
\multicolumn{1}{r}{} &  & \textsf{\scriptsize{}Adaptive} & \textsf{\scriptsize{}Single} & \multicolumn{1}{c}{\textsf{\scriptsize{}Parallel}}\tabularnewline
\cline{2-5} 
\multirow{4}{*}{\begin{turn}{90}
\textsf{\scriptsize{}Leader}
\end{turn}} & \textsf{\scriptsize{}Std. dev.} & \textsf{\scriptsize{}$\mathsf{4.8\times10^{-8}}$} & \textsf{\scriptsize{}$\mathsf{1.3\times10^{-7}}$} & \textsf{\scriptsize{}$\mathsf{5.2\times10^{-8}}$}\tabularnewline
 & \textsf{\scriptsize{}Levels} & \textsf{\scriptsize{}20} & \textsf{\scriptsize{}6} & \textsf{\scriptsize{}6}\tabularnewline
 & \textsf{\scriptsize{}Budget} & \textsf{\scriptsize{}1000} & \textsf{\scriptsize{}1000} & \textsf{\scriptsize{}$\mathsf{5\times1000}$}\tabularnewline
 & \textsf{\scriptsize{}Time (MC)} & \textsf{\scriptsize{}7.3s (30h)} & \textsf{\scriptsize{}2.5s (4.4h)} & \textsf{\scriptsize{}5.8s (5.0h)}\tabularnewline
\cline{2-5} 
\multirow{4}{*}{\begin{turn}{90}
\textsf{\scriptsize{}Philosophers}
\end{turn}} & \textsf{\scriptsize{}Std. dev.} & \textsf{\scriptsize{}$\mathsf{4.2\times10^{-7}}$} & \textsf{\scriptsize{}$\mathsf{7.7\times10^{-7}}$} & \textsf{\scriptsize{}$\mathsf{2.8\times10^{-7}}$}\tabularnewline
 & \textsf{\scriptsize{}Levels} & \textsf{\scriptsize{}109} & \textsf{\scriptsize{}5} & \textsf{\scriptsize{}5}\tabularnewline
 & \textsf{\scriptsize{}Budget} & \textsf{\scriptsize{}1000} & \textsf{\scriptsize{}1000} & \textsf{\scriptsize{}$\mathsf{5\times1000}$}\tabularnewline
 & \textsf{\scriptsize{}Time (MC)} & \textsf{\scriptsize{}5.4s (2.3h)} & \textsf{\scriptsize{}1.7s (41m)} & \textsf{\scriptsize{}3.7s (1.4h)}\tabularnewline
\cline{2-5} 
\multirow{4}{*}{\begin{turn}{90}
\textsf{\scriptsize{}Counters}
\end{turn}} & \textsf{\scriptsize{}Std. dev.} & \textsf{\scriptsize{}$\mathsf{2.1\times10^{-7}}$} & \textsf{\scriptsize{}$\mathsf{5.0\times10^{-7}}$} & \textsf{\scriptsize{}$\mathsf{2.3\times10^{-7}}$}\tabularnewline
 & \textsf{\scriptsize{}Levels} & \textsf{\scriptsize{}3942} & \textsf{\scriptsize{}4} & \textsf{\scriptsize{}4}\tabularnewline
 & \textsf{\scriptsize{}Budget} & \textsf{\scriptsize{}500} & \textsf{\scriptsize{}500} & \textsf{\scriptsize{}$\mathsf{5\times500}$}\tabularnewline
 & \textsf{\scriptsize{}Time (MC)} & \textsf{\scriptsize{}15s (7.5h)} & \textsf{\scriptsize{}2.8s (1.2h)} & \textsf{\scriptsize{}4.8s (1.9h)}\tabularnewline
\cline{2-5} 
\end{tabular}{\scriptsize \par}

\captionof{table}{Summary of results.\label{tab:summary}}%
\end{minipage}
\end{figure}

For each model we performed a number of experiments to compare the
performance of the fixed and adaptive importance splitting algorithms
with and without distribution, using different budgets and levels.
Our results are illustrated in the form of empirical cumulative probability
distributions of $100$ estimates, noting that a perfect (zero variance)
estimator distribution would be a single step. The results are also
summarised in Table \ref{tab:summary}. The probabilities we estimate
are all close to $10^{-6}$ and are marked on the figures with a vertical
line. Since we are not able to use numerical techniques to calculate
the true probabilities, we use the average of $200$ low variance
estimates as our best overall estimate.

As a reference, we applied the adaptive algorithm to each model using
a single computational thread. We chose parameters to maximise the
number of levels and thus minimise the variance for a given score
function and budget. The resulting distributions, sampled at every
tenth percentile, are plotted with circular markers in the figures.
Over these points we superimpose the results of applying a single
instance of the fixed level algorithm with just a few levels. We also
superimpose the average estimates of five parallel threads running
the fixed level algorithm, using the same levels.

The figures confirm our expectation that the fixed level algorithm
with few levels is outperformed by the adaptive algorithm. The figures
also demonstrate that the average of parallel instances of the fixed
level algorithm are very close to the performance of the adaptive
algorithm. The timings given in Table \ref{tab:summary} show that
the distributed approach achieves these results in less time. For
comparison we also include the estimated time of using a simple Monte
Carlo (MC) estimator to achieve the same standard deviation. Importance
splitting gives more than three orders of magnitude improvement in
all cases. All results were generated using an Intel Core i7-3740
CPU with 4 cores running at 2.7 GHz.

In the remainder of this section we briefly describe our models and
their associated properties and score functions.

\begin{figure}
\begin{minipage}[t]{0.48\columnwidth}%
\includegraphics[width=0.9\linewidth]{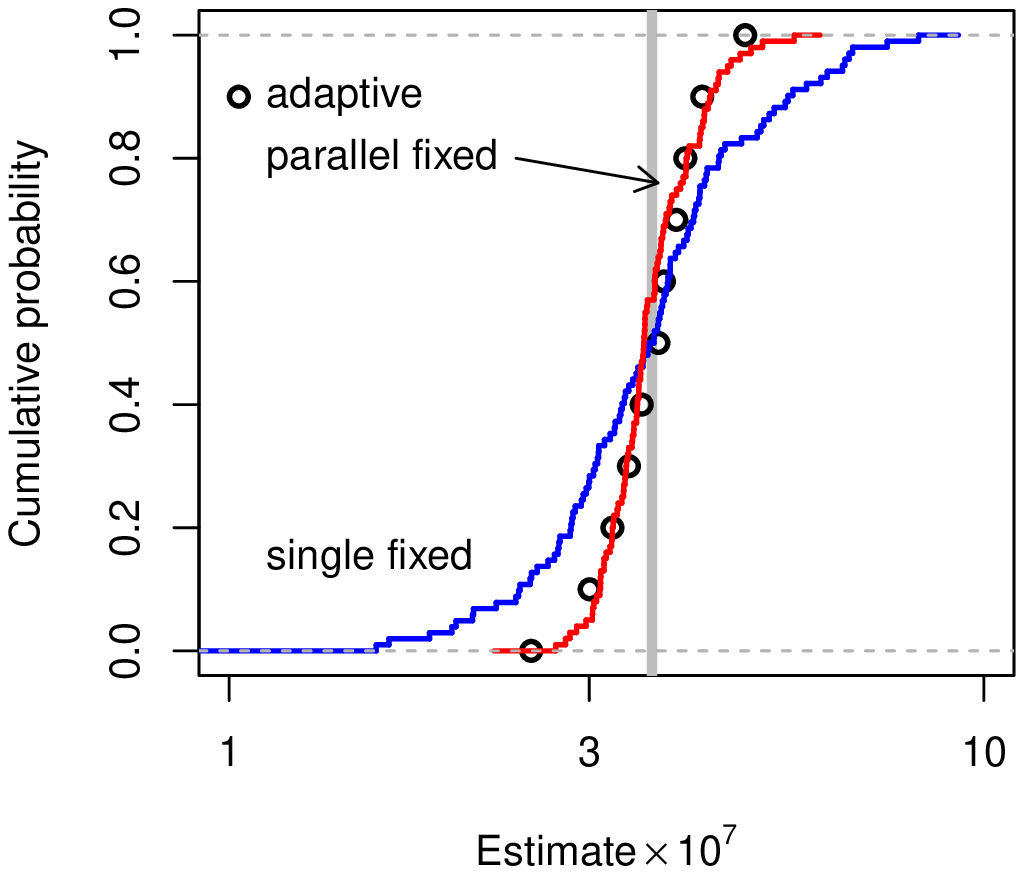}\protect\caption{Leader election.\label{fig:leader}}
\end{minipage}\quad{}%
\begin{minipage}[t]{0.48\columnwidth}%
\includegraphics[width=0.9\textwidth]{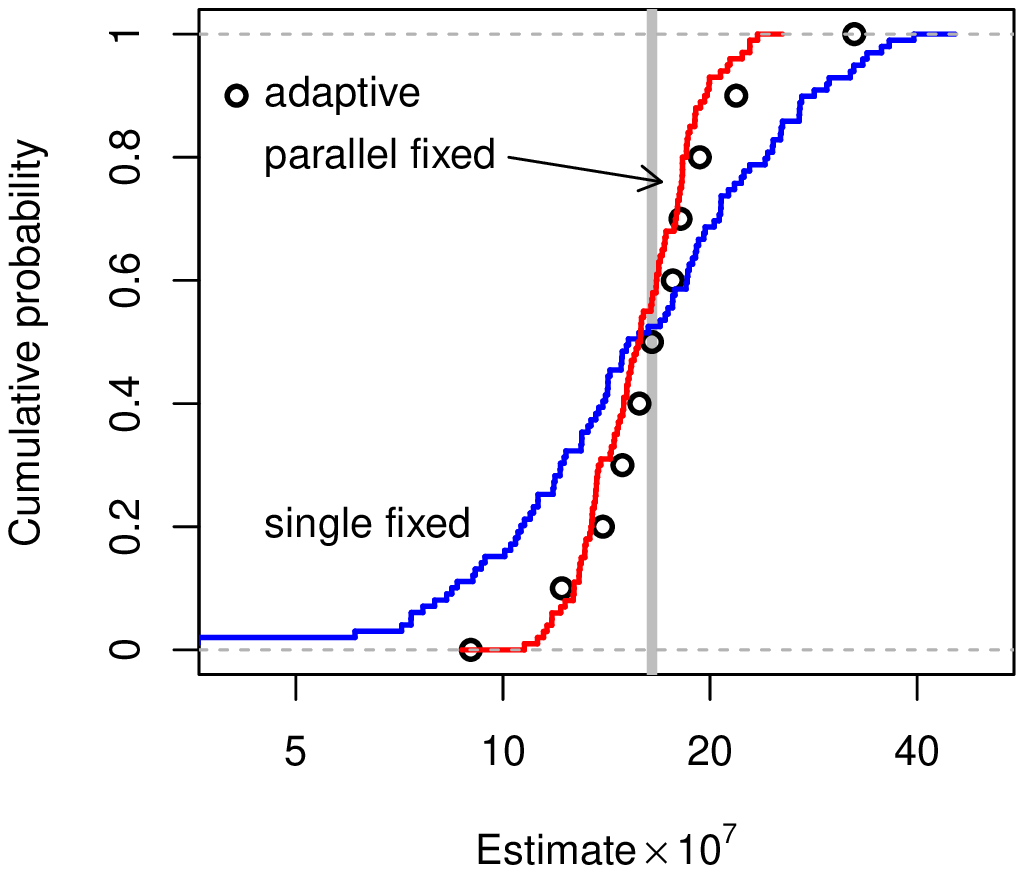}\protect\caption{Dining philosophers.\label{fig:philosophers}}
\end{minipage}
\end{figure}

\subsubsection*{Leader Election}

Our leader election case study is based on the \textsc{Prism} model
of the synchronous leader election protocol of \cite{ItaiRodeh1990}.
With $N=20$ processes and $K=6$ probabilistic choices the model
has approximately $1.2\times10^{18}$ states. We consider the probability
of the property $\mathbf{G}^{420}\neg\mathit{elected}$, where $\mathit{elected}$
denotes the state where a leader has been elected. Our chosen score
function uses the time bound of the $\mathbf{G}$ operator to give
nominal scores between $0$ and $420$. The model constrains these
to only $20$ actual levels, but with evenly distributed probability.
For the fixed level algorithm we use scores of $70,140,210,280,350$
and $420$.

\subsubsection*{Dining Philosophers}

Our dining philosophers case study extends the Prism model of the
fair probabilistic protocol of \cite{LehmannRabin1981}. With 150
philosophers our model contains approximately $2.3\times10^{144}$
states. We consider the probability of the property $\mathbf{F}^{30}\textit{Phil eats}$,
where $\mathit{Phil}$ is the name of an arbitrary philosopher. The
adaptive algorithm uses the heuristic score function described in
\cite{JegourelLegaySedwards2014}, which includes the five logical
levels used by the fixed level algorithm. The heuristic favours short
paths, based on the assumption that as time runs out the property
is less likely to be satisfied.

\vspace{-1em}

\subsubsection*{Dependent Counters}

Our dependent counters case study comprises ten counters, initially
set to zero, that with some probability dependent on the values of
the other counters are either incremented or reset to zero. This can
be viewed as modelling an abstract computational process, a set of
reservoirs of finite capacity, or as the failure and repair of ten
different types of components in a system, etc. With a maximum count
of $10$, the model has approximately $2.6\times10^{10}$ states.

We consider the probability of the property $\mathbf{X}^{1}(\neg\mathit{init}\mathbf{U}^{1000}\mathit{complete})$,
where $\mathit{init}$ and $\mathit{complete}$ denote the initial
state and the state where all counters have reached their maximum
value. Our score function ranges over values between 0 and 99, but
the probabilities are not evenly distributed. With a budget of $500$,
uniformly distributed fixed scores fail to produce traces that satisfy
the property until the difference between the last two levels is about
$5$. Note that our budget is limited to only 500 simulations due
to the length of the traces that must be stored by the adaptive algorithm.
We maintain this budget for the fixed level algorithm to simplify
comparison. After a small amount of trial and error, we adopted fixed
scores of $80,90,95$ and $99$.

\section{Challenges and Prospects}

Our results demonstrate the effectiveness and flexibility of our framework
with discrete time properties applied to standard case studies. Future
challenges include industrial scale examples and the implementation
of continuous time properties. We also intend to provide proofs of
the correctness of our observers and of our logic's memory requirements.

Although the manual construction of score functions adds to the overall
cost of using importance splitting, we believe that distribution relaxes
the need for these to be highly optimised. We also expect that it
will be possible to construct good score functions automatically using
statistical learning techniques.

\bibliographystyle{abbrv}
\bibliography{LightweightObservers}

\begin{thebibliography}{10}

\bibitem{AlurHenzinger1999}
R.~Alur and T.~A. Henzinger.
\newblock Reactive modules.
\newblock {\em Formal Methods in System Design}, 15(1):7--48, 1999.

\bibitem{BauerLeuckerSchallhart2006}
A.~Bauer, M.~Leucker, and C.~Schallhart.
\newblock Monitoring of real-time properties.
\newblock In {\em FSTTCS 2006: Foundations of Software Technology and
  Theoretical Computer Science}, pages 260--272. Springer, 2006.

\bibitem{BoyerCorreLegaySedwards2013}
B.~Boyer, K.~Corre, A.~Legay, and S.~Sedwards.
\newblock {PLASMA-lab}: A flexible, distributable statistical model checking
  library.
\newblock In K.~Joshi, M.~Siegle, M.~Stoelinga, and P.~R. D'Argenio, editors,
  {\em Quantitative Evaluation of Systems}, volume 8054 of {\em LNCS}, pages
  160--164. Springer, 2013.

\bibitem{CerouDelMoralFuronGuyader2012}
F.~C{\'e}rou, P.~Del~Moral, T.~Furon, and A.~Guyader.
\newblock Sequential {M}onte {C}arlo for rare event estimation.
\newblock {\em Statistics and Computing}, 22:795--808, 2012.

\bibitem{CerouGuyader2007}
F.~C{\'e}rou and A.~Guyader.
\newblock Adaptive multilevel splitting for rare event analysis.
\newblock {\em Stochastic Analysis and Applications}, 25:417--443, 2007.

\bibitem{ClarkeEmersonAllenSifakis2009}
E.~M. Clarke, E.~A. Emerson, and J.~Sifakis.
\newblock Model checking: algorithmic verification and debugging.
\newblock {\em Commun. ACM}, 52(11):74--84, November 2009.

\bibitem{Dijkstra1975}
E.~W. Dijkstra.
\newblock Guarded commands, nondeterminacy and formal derivation of programs.
\newblock {\em Commun. ACM}, 18(8):453--457, August 1975.

\bibitem{Eisner-et-al2003}
C.~Eisner, D.~Fisman, J.~Havlicek, Y.~Lustig, A.~McIsaac, and
  D.~Van~Campenhout.
\newblock Reasoning with temporal logic on truncated paths.
\newblock In {\em Computer Aided Verification}, pages 27--39. Springer, 2003.

\bibitem{FinkbeinerSipma2004}
B.~Finkbeiner and H.~Sipma.
\newblock Checking finite traces using alternating automata.
\newblock {\em Formal Methods in System Design}, 24(2):101--127, 2004.

\bibitem{Geilen2001}
M.~C.~W. Geilen.
\newblock On the construction of monitors for temporal logic properties.
\newblock {\em Electronic Notes in Theoretical Computer Science},
  55(2):181--199, 2001.

\bibitem{GerthPeledVardiWolper1995}
R.~Gerth, D.~Peled, M.~Y.~V. Vardi, and P.~Wolper.
\newblock Simple on-the-fly automatic verification of linear temporal logic.
\newblock In {\em In Protocol Specification Testing and Verification}, pages
  3--18. Chapman \& Hall, 1995.

\bibitem{GiannakopoulouHavelund2001}
D.~Giannakopoulou and K.~Havelund.
\newblock Automata-based verification of temporal properties on running
  programs.
\newblock In {\em Proceedings of 16th Annual International Conference on
  Automated Software Engineering}, pages 412--416. IEEE, Nov 2001.

\bibitem{HammersleyHandscomb1964}
J.~M. Hammersley and D.~C. Handscomb.
\newblock {\em {M}onte {C}arlo Methods}.
\newblock Methuen \& Co., 1964.

\bibitem{HavelundRosu2002}
K.~Havelund and G.~Ro\c{s}u.
\newblock Synthesizing monitors for safety properties.
\newblock In J.-P. Katoen and P.~Stevens, editors, {\em Tools and Algorithms
  for the Construction and Analysis of Systems}, volume 2280 of {\em Lecture
  Notes in Computer Science}, pages 342--356. Springer, 2002.

\bibitem{ItaiRodeh1990}
A.~Itai and M.~Rodeh.
\newblock Symmetry breaking in distributed networks.
\newblock {\em Information and Computation}, 88(1):60--87, 1990.

\bibitem{JegourelLegaySedwards2012}
C.~Jegourel, A.~Legay, and S.~Sedwards.
\newblock Cross-entropy optimisation of importance sampling parameters for
  statistical model checking.
\newblock In P.~Madhusudan and S.~A. Seshia, editors, {\em Computer Aided
  Verification}, volume 7358 of {\em LNCS}, pages 327--342. Springer, 2012.

\bibitem{SedwardsJegourelLegay2012}
C.~Jegourel, A.~Legay, and S.~Sedwards.
\newblock A platform for high performance statistical model checking --
  {PLASMA}.
\newblock In C.~Flanagan and B.~K\"onig, editors, {\em Tools and Algorithms for
  the Construction and Analysis of Systems}, volume 7214 of {\em LNCS}, pages
  498--503. Springer, 2012.

\bibitem{JegourelLegaySedwards2013}
C.~Jegourel, A.~Legay, and S.~Sedwards.
\newblock Importance splitting for statistical model checking rare properties.
\newblock In {\em Computer Aided Verification}, volume 8044 of {\em LNCS},
  pages 576--591. Springer, 2013.

\bibitem{JegourelLegaySedwards2014}
C.~Jegourel, A.~Legay, and S.~Sedwards.
\newblock An effective heuristic for adaptive importance splitting in
  statistical model checking.
\newblock In T.~Margaria and B.~Steffen, editors, {\em Leveraging Applications
  of Formal Methods, Verification and Validation. Specialized Techniques and
  Applications}, volume 8803 of {\em LNCS}, pages 143--159. Springer, 2014.

\bibitem{Kahn1950}
H.~Kahn.
\newblock Random sampling ({M}onte {C}arlo) techniques in neutron attenuation
  problems.
\newblock {\em Nucleonics}, 6(5):27, 1950.

\bibitem{KahnHarris1951}
H.~Kahn and T.~E. Harris.
\newblock Estimation of particle transmission by random sampling.
\newblock In {\em Applied Mathematics}, volume~5 of {\em series 12}. National
  Bureau of Standards, 1951.

\bibitem{KahnMarshall1953}
H.~Kahn and A.~W. Marshall.
\newblock Methods of reducing sample size in {M}onte {C}arlo computations.
\newblock {\em Operations Research}, 1(5):263--278, November 1953.

\bibitem{LehmannRabin1981}
D.~Lehmann and M.~O. Rabin.
\newblock On the advantage of free choice: A symmetric and fully distributed
  solution to the dining philosophers problem.
\newblock In {\em Proc. $8^\mathit{th}$Ann. Symposium on Principles of
  Programming Languages}, pages 133--138, 1981.

\bibitem{MetropolisUlam1949}
N.~Metropolis and S.~Ulam.
\newblock The {M}onte {C}arlo method.
\newblock {\em Journal of the American Statistical Association},
  44(247):335--341, September 1949.

\bibitem{Niederreiter1992}
H.~Niederreiter.
\newblock {\em Random Number Generation and Quasi-{M}onte {C}arlo Methods}.
\newblock Society for Industrial and Applied Mathematics, 1992.

\bibitem{RubinoTuffin2009}
G.~Rubino and B.~Tufin~(eds.).
\newblock {\em Rare Event Simulation using {M}onte {C}arlo Methods}.
\newblock John Wiley \& Sons, Ltd, 2009.

\bibitem{Villen-Altamirano1991}
M.~Vill{\'e}n-Altamirano and J.~Vill{\'e}n-Altamirano.
\newblock {RESTART}: A method for accelerating rare event simulations.
\newblock In J.~W. Cohen and C.~D. Pack, editors, {\em Queueing, Performance
  and Control in ATM}, pages 71--76. Elsevier, 1991.

\bibitem{YounesKwiatkowskaNormanParker2006}
H.~S. Younes, M.~Kwiatkowska, G.~Norman, and D.~Parker.
\newblock Numerical vs. statistical probabilistic model checking.
\newblock {\em International Journal on Software Tools for Technology
  Transfer}, 8(3):216--228, 2006.

\end{thebibliography}

\end{document}